\documentclass{emulateapj}

\slugcomment{accepted November 7, 2005, for publication in ApJ Letters}
\shorttitle{Dust surrounding HD 32297}
\shortauthors{P. Kalas}

\begin{document}


\title{First optical images of circumstellar dust surrounding the debris disk candidate HD 32297}

\author{Paul Kalas\altaffilmark{1,2}}
\affil{Astronomy Department, 
University of California, Berkeley, CA 94720}
\email{kalas@astron.berkeley.edu}

\altaffiltext{1}{Astronomy Department, 
University of California, Berkeley, CA 94720}
\altaffiltext{2}{National Science Foundation Center for Adaptive Optics, University of California, Santa Cruz, CA 95064}

\begin{abstract}
Near-infrared imaging with the {\it Hubble Space Telescope} recently revealed a
circumstellar dust disk around the A star HD 32297.  Dust scattered 
light is detected as far as 400 AU radius and the linear morphology is consistent with
a disk $\sim$10$\degr$ away from an edge-on orientation.  Here we present the first optical
images that show the dust scattered light morphology from 560 to 1680 AU radius.  The position angle of the
putative disk midplane diverges by $\sim$31$\degr$ and
the color of dust scattering is most likely blue.  We associate
HD 32297 with a wall of interstellar gas and the enigmatic region south of the Taurus molecular cloud.  
We propose that the extreme asymmetries and
blue disk color originate
from a collision with a clump of interstellar material as HD 32297 moves southward, and
discuss evidence consistent with an age of 30 Myr or younger.
\end{abstract}
,

\section{Introduction}
Debris disks are the exosolar analogs of our Zodiacal light and Kuiper Belt and
each new discovery represents an opportunity to understand how planetary
systems form and evolve around other stars.  
Schneider, Silverstone \& Hines (2005) recently showed
that HD 32297, an A star at $\sim$112$^{+15}_{-12}$ pc, illuminates
a dusty nebulosity resembling the edge-on debris
disks around $\beta$ Pic \citep{smith84} and AU Mic \citep{kal04}.
HD 32297 was one of 26 stars that they identified as
debris disk candidates for coronagraphic imaging with the
NICMOS camera aboard  the {\it Hubble Space Telescope}.  
Using the F110W filter ($\lambda_c$= 1104 nm,
$\Delta\lambda$ = 592 nm), the HD 32297 disk was found to be extended by at least 400 AU (3.3$\arcsec$)
to the northeast with PA = 47.6$\pm1\degr$ (G. Schneider, 2005, private communication), 
and at least 250 AU to the southwest.  The hundreds of AU extent
of the disk, and the significant asymmetry, are indeed comparable to
those of $\beta$ Pic \citep{kal95}.  
Here we present new $R$-band observations of HD 32297 using a ground-based
coronagraphic camera that reveal a larger and more asymmetric circumstellar nebulosity than
shown by the HST data.

\section{Observations \& Data Analysis}

We artificially eclipsed HD 32297 using
an optical stellar coronagraph at the University of Hawaii 2.2-m
telescope on Mauna Kea, Hawaii \citep{kal96a}.  
Data were acquired with a Tek
2048$\times$2048 CCD with a scale of 0.407$\arcsec$/pixel and through
a standard broadband $R$ filter ($\lambda_c$=647 nm,
$\Delta\lambda$ = 125 nm).  
Observations were made on 28 September, 2005, with a 6.5$\arcsec$ diameter occulting spot
and 1320 seconds effective integration time.
Measurements of photometric standard stars showed photometric condition, with image quality, as measured by the full-width at half-maximum (FWHM)
of field stars, equal to $\sim$1.2$\arcsec$.  A series of short, unocculted integrations 
yielded $R = 7.9 \pm 0.1$ mag for HD 32297.
We also observed three other bright stars with the coronagraph to
check for spurious features such as diffraction spikes and internal reflections.

After the data were bias subtracted, flat-fielded
and sky-subtracted, we subtracted the stellar
point spread function (PSF) to remove
excess stellar light from around the occulting spot.  We used the real PSF's
from other stars observed throughout each night, as well as artificial PSF's.
Artificial PSF subtraction is effective for HD 32297 because the circumstellar disk is close to
edge-on.  We extracted the stellar PSF for each image of HD 32297
by sampling the image radially in
a direction perpendicular to the PA of the disk.
We then fit a polynomial to the data and generated an 
artificial PSF that is a figure of rotation of the polynomial.  The PSF's were then scaled
and registered to each data frame such that subtraction minimized 
the residual light in directions perpendicular to the disk beyond the edge of the
occulting spot. 

\section{Results}

Figure 1 presents our $R$-band image of nebulosity surrounding HD 32297.  The inner detection
limit is 5.0$\arcsec$ and the nebulosity is detected as far as 15$\arcsec$ (1680 AU) radius.  On these
spatial scales the two ansae taken together do not resemble a circumstellar disk because the apparent
midplanes diverge in position angle.  Instead the curved morphology resembles that of pre-main sequence
stars such as SU Aur and Z CMa \citep{nakajima95}.  The northeast side is a relatively narrow structure
resembling the near edge-on disk described by \citet{ssh05}, but with PA = 34$\pm$1$\degr$ that is
13.6$\degr$ smaller than that measured in the HST NICMOS data.  The southwest side of the nebulosity
is a broader structure that curves westward with radius.  We adopt PA = $245\pm2\degr$, which is 18$\degr$ away
from the midplane PA measured by \citet{ssh05}, and forms a 31$\degr$ angle with the northeast midplane in our data.  
 The FWHM of the disk perpendicular to the midplane at $8\arcsec$ radius is $3.7\arcsec$ and $5.0\arcsec$ for
 the NE and SW sides, respectively.

The midplane radial surface brightness profiles for the SW side is $0.3-0.5$ mag arcsec$^{-2}$ brighter than
the NE side between 5$\arcsec - 10\arcsec$ radius, and approximately equal further out (Fig. 2).  
PSF subtraction dominates the uncertainty in the absolute flux measurements, but the relative flux measurements
between the NE and SW sides remain constant between different PSF subtractions.  
The cumulative magnitudes for the NE and SW extensions are equal to within a
tenth of a magnitude, with $R$=$20.0\pm0.5$ mag for each side.  Again the uncertainty
depends on the PSF subtraction and scales upward or downward uniformly for both
sides of the disk.  To first order, both extensions have the same scattering cross-section of dust, even though the spatial
distribution is significantly different, similar to the findings for the $\beta$ Pic disk \citep{kal95}.

Between 5$\arcsec - 15\arcsec$ radius the two midplane profiles can be described by power-laws
with indices $-2.7\pm0.2$ and $-3.1\pm0.2$ for the NE and SW sides, respectively (Fig. 2).  
These indices are comparable to the $F110W$ surface brightness profile of the NE midplane
between 1.6$\arcsec$ and 3.3$\arcsec$ radius \citep{ssh05}.  
The similarity supports extrapolating the $R$-band surface brightness profile inward 
to estimate the $R - F110W$ disk color within 3.3$\arcsec$ radius (Fig. 2).  We find $R - F110W \approx -1$ mag for the
NE extension and $-2$ mag for the SW extension, whereas
the intrinsic stellar color is $R - F110W$ = +0.21 mag.  The blue scattered light color is consistent with sub-micron
Raleigh scattering grains found in the interstellar medium \citep{draine03}, as well as the
outer region of the AU Mic debris disk \citep{metchev05}.  If HD 32297 is comparable in spectral
type to $\beta$ Pic (A5V), then grains with radii smaller than $\sim$5 $\mu$m will be blown 
out of the system on one dynamical timescale ($\sim$10$^3$ yr;  Artymowicz \& Clampin 1997).
Below we discuss how the presence of small grains ($\sim$0.1 $\mu$m) leads to several
plausible scenarios for the origin of the nebulosity and the age of the system.

\section{Discussion}

The asymmetric, large-scale morphology and the blue color of nebulosity surrounding HD 32297 indicate
that a population of dust grains may be primordial originating from the interstellar medium.  Interstellar
grains have a size distribution that peaks at $0.1-0.2$ $\mu$m \citep{kim94, mathis96} and many
reflection nebulosities have a blue color \citep{witt86}.
However, the morphology of the HD 32297 nebulosity
between 0.5$\arcsec$ and 1.7$\arcsec$ radius satisfies 
four criteria for the imaging detection of a circumstellar disk \citep{kal96b}.
In this inner region the disk is relatively symmetric and a power-law fit to the surface 
brightness profile has index -3.6 \citep{ssh05}, which
is comparable to the outer disk regions of $\beta$ Pic and
AU Mic \citep{kal04}.  The steepness of this surface brightness profile is consistent with models of an
outward propagation of grains from an interior source region due to radiation pressure \citep{aug01}.
Beyond 1.7$\arcsec$ (190 AU) radius the disk may overlap with an interstellar nebulosity
or it is influenced by forces that are otherwise insignificant in the inner disk.

If the large-scale nebulosity is produced by
a random encounter between an A star and a clump of interstellar gas and dust, then the resulting morphology
should demonstrate the signature linear filamentary features of the Pleiades Phenomenon \citep{kal02}.
We do not detect Pleiades-like nebulosity, though
interaction with the ISM is nevertheless plausible as the galactic location of HD 32297 ($l$=192.83$\degr$, $b$=-20.17$\degr$) 
coincides with a ridge of relatively high density gas 
outside of our local bubble (Fig 3; Kalas et al. 2002).  This ridge also contains two 
stars surrounded by optical nebulosity that are members of the Pleiades open cluster
(M45; d = $118\pm4$ pc; van Leeuwen 1999). 
The proper motion vector of HD 32297 
($\mu_{\alpha}=7$ mas/yr, $\mu_{\delta}=-20$ mas/yr) points to the south-southeast, with a sky-plane motion
corresponding to 13.4 km s$^{-1}$ at 112 pc distance.  Therefore the southern side of the
disk will suffer enhanced erosion that would result in both a brighter nebulosity and diminished
disk mass, compared to the northern side of the disk.  \citet{liss89} refer to this process
as ISM sandblasting, and \citet{art97} show that stellar radiation pressure would protect the circumstellar
disk from the ISM up to a few hundred AU radius from the star.  The ISM avoidance radius is a
function of several factors, such as ISM density, relative velocity, and encounter geometry.  
A more detailed model applied specifically to HD 32297 is required to understand if the observed
disk asymmetries are consistent with ISM sandblasting.  However, \citet{art97} cautioned
that ISM grains do not have sufficient mass
to perturb grains vertically away from a disk  midplane.  If this is valid, then other processes
could create the observed $R$-band asymmetries,
such as the entrainment of small grains by the ISM gas that should be associated with the ISM dust, or
dynamical perturbations from the two stars, HD 32304 (G5, $d= 134_{-15}^{+18}$ pc)
and BD +7 777s, south-southeast of HD 32297 (Figs. 1 \& 4).

In the ISM sandblasting scenario, HD 32297 could be a main sequence star presently
undergoing a random encounter with a clump of ISM.  An alternate hypothesis is that HD 32297 is very young,
and the nebulosity resembles that of SU Aur and Z CMa because the dust is the remnant
of an outflow cavity, or more generally represents pristine matter from the natal cloud.  The position angle
discrepancies could arise because HST NICMOS is sensitive to the circumstellar disk at $<$200 AU radius with
PA $\approx48\degr$, whereas
our $R$ band data observe the top (out of the sky-plane) of an outflow cavity associated with this inner disk.  
However, the lack of reddening ($V - K$ = 0.54 mag, $J - H$ = 0.06, $H - K$ = 0.03) generally argues 
against a massive obscuring dust disk, such as that discovered around the Herbig Ae/Be 
Star PDS 144N \citep{perrin05}.  
Examination of several degrees of sky surrounding HD 32297 in the Digitized Sky Survey
reveals filamentary nebulosities  to the southeast and southwest, apparently 
associated with the $\lambda$ Orionis molecular ring (SH 2-264) to the east  (Fig. 4).  
$\lambda$ Orionis is in the background at $\sim$400 pc, and its diameter is no
greater than 20 pc in radius \citep{madd87, dolan02}.   Therefore, HD 32297 must
originate from a different star forming region.
Taurus-Aurigae, $\sim$10$\degr$ to the north of HD 32297 and with heliocentric distance $140\pm20$ pc \citep{elias78},
may be a possibility.  The proper motion of HD 32297 is comparable to
T Tauri stars in this region \citep{frink97}, but it does not have a relative excess southward 
that would flag HD 32297 as a runaway star.   Therefore, if HD 32297 is indeed a pre-main sequence
star, then it formed in relative isolation from Taurus-Aurigae within an outlying clump of the
main molecular cloud.   

The discovery of young stars south of Taurus led \citet{neu97} through
a similar  considerations when discussing the 
origin of their lithium-rich targets.  In addition, they proposed two alternatives
that may be applicable to HD 32297.  First, they noted that the midplane
of the Gould Belt, with age $\sim$30 Myr, passes south of 
Taurus.  In fact, HD 32297 is located within the Gould Belt midplane.  
Second, star formation in Taurus-Aurigae may have
been triggered 30 Myr ago when a high-velocity cloud passed northward
through the galactic plane \citep{lepine94}.  The first generation of stars formed
at this earlier epoch
within the molecular cloud, which eventually passed again southward through the 
galactic plane.  The second passage formed a new generation of stars observed
today in Taurus-Aurigae, with the first generation separating from the natal material
when the latter decelerated as it encountered denser ISM in the galactic plane.  The first-generation stars 
are currently found south of Taurus (both in galactic latitude and declination).  
Therefore, whether associated with the Gould Belt or with Taurus-Aurigae,
the young stars presently located south of the Taurus molecular cloud have age $\sim$30 Myr.
HD 32297 may be a member of either group given that the relatively large dust disk
surrounding it may be adopted as a proxy for spectroscopic youth indicators.

\section{Conclusions and Future Work}

Coronagraphic $R$-band data reveal that the circumstellar nebulosity surrounding HD 32297
is significantly distorted relative to the near edge-on disk observed within 400 AU with
HST NICMOS.  We detect nebulosity as far as 1680 AU radius, with long axes that deviate
from the NICMOS position angles by $\sim$15$\degr$ for each midplane.  The southwest midplane
is warped and vertically distended.   We invoke the possibility of ISM sandblasting, which is consistent
with the southward proper motion of HD 32297.  We examine several scenarios relating
to the age and origin of HD 32297.  Association with either the Gould Belt or Taurus-Aurigae
would give age $\sim$30 Myr, similar to that of $\beta$ Pic.  A younger age is also possible if 
HD 32297 formed in an isolated, outlying cloud of Taurus-Aurigae.  

If the ISM sandblasting scenario is correct, then future multi-color imaging should
reveal significant color differences between the northeast and southwest sides
of the disk.  The southwest side may also show significant color structure perpendicular
to disk midplane if the apparent distortions in the disk are due to small grains swept 
northward.  Dynamical perturbations are possible if either HD 32304 or BD +7 777s to
the southeast are physically associated with HD 32297.
Future spectroscopic observations of HD 32297 should also include these two stars to constrain
their spectral types, relative motions, and ages.  If the radial velocities support a physical
association, then either HD 32304 or BD +7 777s may demonstrate additional age
indicators that would constrain the evolutionary status of HD 32297.  Moreover, if HD 32297
is a $\beta$ Pic analog, then multi-epoch spectroscopy should test for the variable, transient
redshifted features thought to arise from cometary activity under the dynamical
influence of planets \citep{beust00}.

\acknowledgements
{\bf Acknowledgements:}  This work was supported in part by NASA grants 
to P.K, and by the NSF Center for Adaptive Optics, managed
by the University of California, Santa Cruz, under cooperative
agreement AST 98-76783.
We thank Michael Ratner (SAO/Harvard) for supporting the observing campaigns
at the University of Hawaii 2.2-m telescope.

\clearpage


\begin{figure}
\epsscale{0.7}
\plotone{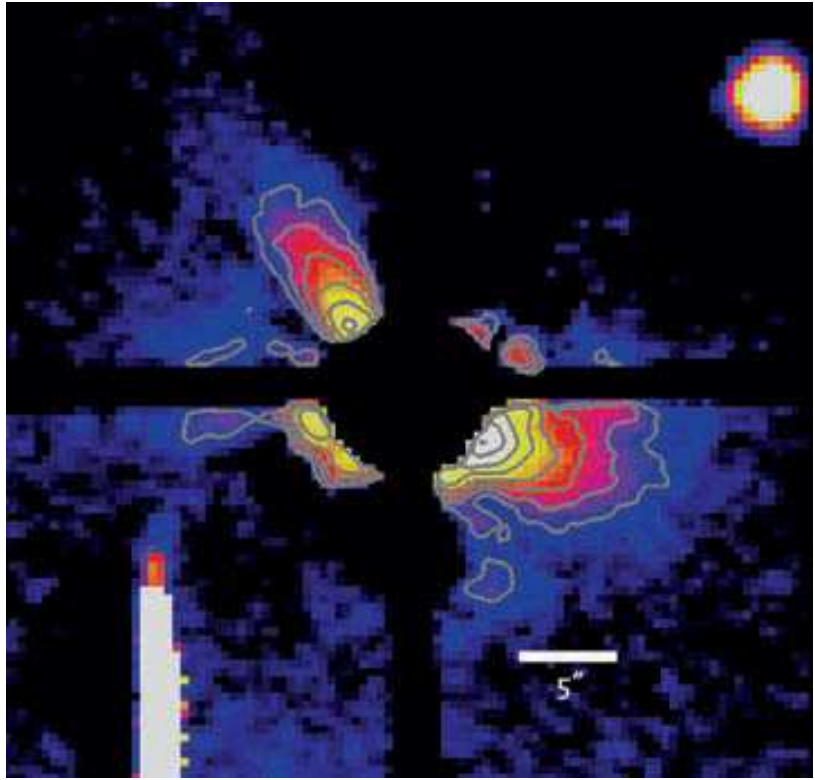}
\caption{
Coronagraphic R-band image of HD 32297 after PSF subtraction.  
Intensity is displayed on a logarithmic scale with false colors.  Contours are spaced at
intervals of 0.5 mag arcsec$^{-2}$, and the outer contour represents
23.5 mag arcsec$^{-2}$.  North is up, east is left, and the field shown is 40$\arcsec\times40\arcsec$.
The bright column from the lower left is a saturation column on the CCD from the star 
BD +7 777s ($V$=10.2,  47.9$\arcsec$ east and 78.6$\arcsec$ south of HD 32297).  
The dominant sources of instrumental scattered light is from the bright star
HD 32304 ($V$=6.9, 61$\arcsec$ east and 134$\arcsec$ south of HD 32297).
\label{fig1}}
\end{figure}


\begin{figure}
\epsscale{1.0}
\plotone{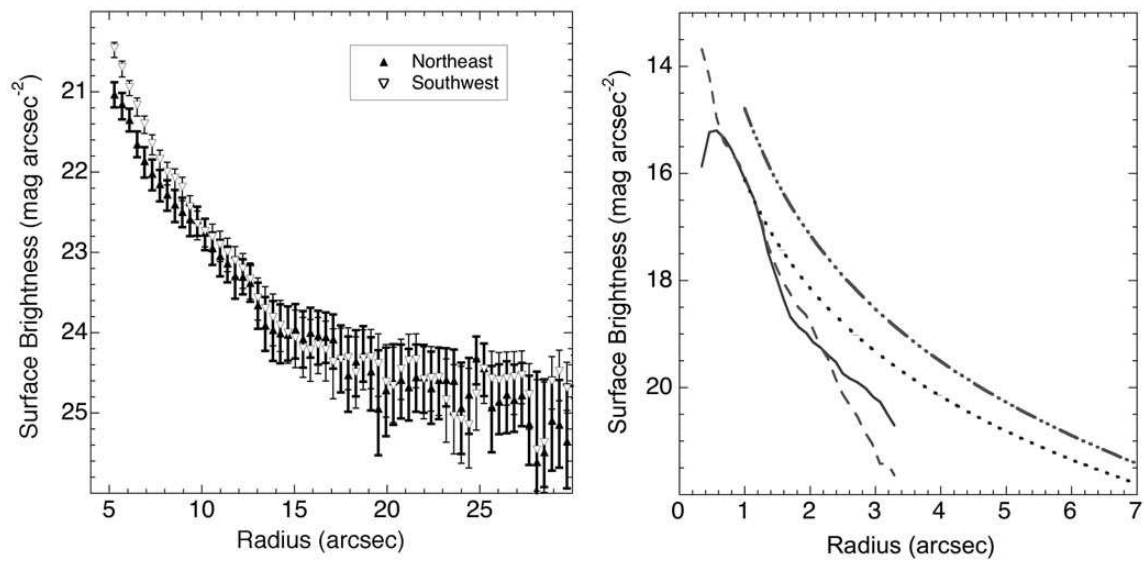}
\caption{
Midplane surface brightness as a function of radius for the northeast and southwest extension of HD 32297.  LEFT:  $R$-band
surface brightness profile of a midplane cut with width 1.2$\arcsec$. Error bars indicate uncertainty from the subtraction of the PSF.
The surface brightness profiles may be fit by power-laws as described in the text.  
RIGHT:  Inward extrapolation of the $R$-band surface brightness profile represented by the power-law fits for
comparison to the HST NICMOS $F110W$ data.  The $R$-band SW disk midplane (dash-dot)
and the NE disk midplane (dot) are both brighter than the $F110W$ midplanes detected by
Schneider et al. 2005 (solid line is the NE extension, dashed line
is the SW extension), consistent with a blue color for the nebulosity.
\label{fig2}}
\end{figure}

\clearpage

\begin{figure}
\epsscale{0.7}
\plotone{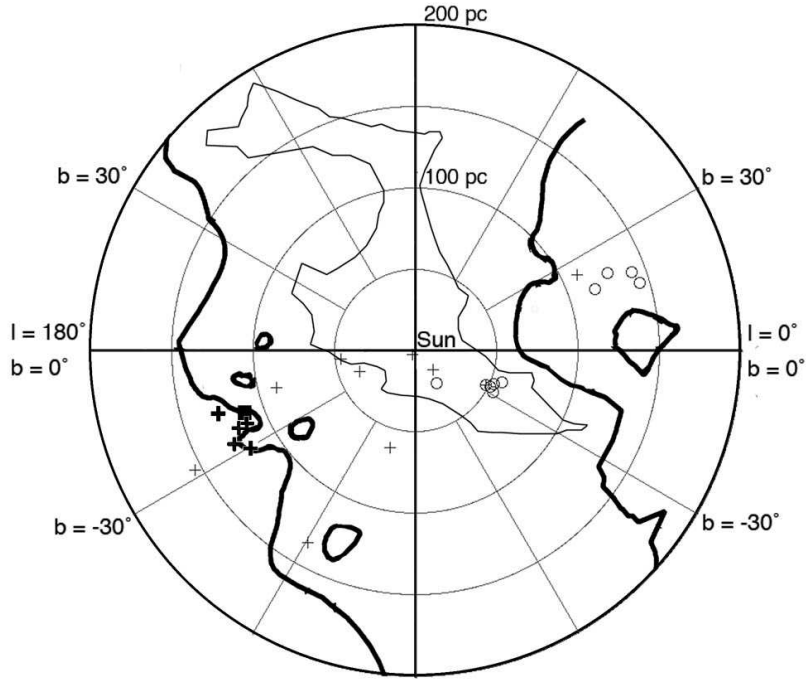}
\caption{
Association of HD 32297 with interstellar gas beyond the local bubble.
HD 32297 is plotted with a solid square (lower left quadrant) in this figure adapted from \citet{kal02} 
showing the spatial relationship
between infrared excess stars and interstellar gas.  This is the meridian plane of the galaxy, containing both
galactic poles and the galactic center to the right.  Thin and thick contours represent neutral gas mapped
by \citet{sfe99} using the Na I D-line doublet.  Other symbols on the map represent the location
of stars that possess the far infrared signature of debris disks.  Stars are mapped here if they 
have $l$=0$\degr\pm$18$\degr$ or $l$=180$\degr\pm$18$\degr$.   HD 32297 is therefore
included in this plot and appears superimposed on a wall or ridge of  relatively
dense interstellar gas ($>50$ m$\AA$ D2-line equivalent width).  Thick crosses mark five more stars that appear to trace the 
boundary of the wall, though the uncertainties in the $Hipparcos$ parallaxes effectively
place them at the same heliocentric distance.
To the left of HD 32297 is HD 28149, and
just below HD 32297 are 18 Tau (HD 23324) and 21 Tau (HD 23432), members of the Pleiades open cluster. 
Two more debris disk candidate stars along the bottom portion of this ridge are HD 28978 and HD 28375
\citep{bap93}.  They lie closest to HD 32297 in galactic latitude.
\label{fig3}}
\end{figure}


\begin{figure}
\epsscale{0.5}
\plotone{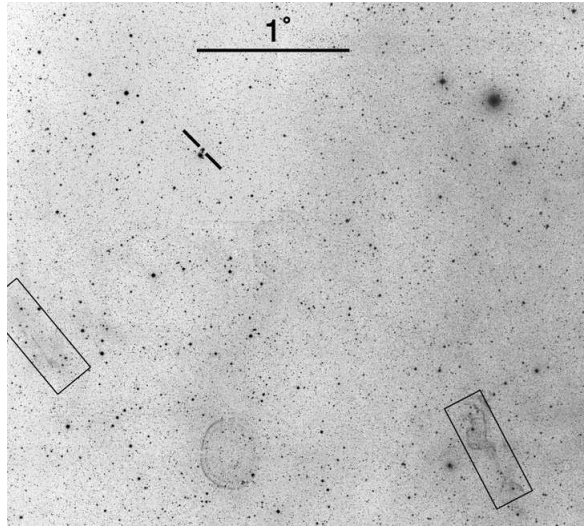}
\caption{
Digitized Sky Survey image of the region around HD 32297 (marked between diagonal lines) shows significant
nebulosity in the large-scale environment (north is up, east is left).  Filamentary H$\alpha$ nebulosities on scales of 
tens of arcminutes (rectangles), are
evident to the southeast (Sh 2-262) and southwest  (Sh 2-260) of HD 32297. They are most likely
associated with $\lambda$ Orionis molecular ring (Sh 2-264) to the east.  The age
and origin of HD 32297 are enigmatic due to the superposition of objects in its environment
that could be associated with $\lambda$ Orionis,
the Gould Belt, or the Taurus molecular cloud.
\label{fig3}}
\end{figure}

\end{document}